\documentclass[review]{elsarticle}

\usepackage{lineno,hyperref}
\modulolinenumbers[5]

\usepackage{natbib}
\usepackage{array}
\usepackage{amsmath}
\usepackage{amsfonts}
\usepackage{multirow}
\usepackage{amsthm}
\usepackage{amssymb,graphicx,tikz,mathabx,amsfonts,mathtools} 
\usepackage{appendix}
\usepackage{booktabs}
\usepackage{algpseudocode}
\usepackage{lipsum}
\usepackage{geometry}
\usepackage{changepage}

\newtheorem{theorem}{Theorem}

\newcommand{\balpha}{\boldsymbol{\alpha}}
\newcommand{\bbeta}{\boldsymbol{\beta}}

\newcommand{\bepsilon}{\boldsymbol{\epsilon}}

\newcommand{\btheta}{\boldsymbol{\theta}}

\newcommand{\bmu}{\boldsymbol{\mu}}

\newcommand{\bxi}{\boldsymbol{\xi}}
\newcommand{\0}{\boldsymbol{0}}

\newcommand{\brho}{\boldsymbol{\rho}}

\newcommand{\bSigma}{\boldsymbol{\Sigma}}

\newcommand{\bOmega}{\boldsymbol{\Omega}}

\newcommand{\bB}{\boldsymbol{B}}

\newcommand{\bA}{\boldsymbol{A}}

\newcommand{\bD}{\boldsymbol{D}}

\newcommand{\bE}{\boldsymbol{E}}

\newcommand{\bH}{\boldsymbol{H}}

\newcommand{\bI}{\boldsymbol{I}}

\newcommand{\bM}{\boldsymbol{M}}

\newcommand{\bP}{\boldsymbol{P}}

\newcommand{\bR}{\boldsymbol{R}}

\newcommand{\bS}{\boldsymbol{S}}
\newcommand{\bt}{\boldsymbol{t}}

\newcommand{\bu}{\boldsymbol{u}}
\newcommand{\bU}{\boldsymbol{U}}

\newcommand{\bV}{\boldsymbol{V}}

\newcommand{\bW}{\boldsymbol{W}}

\newcommand{\bX}{\boldsymbol{X}}
\newcommand{\by}{\boldsymbol{y}}

\newcommand{\bZ}{\boldsymbol{Z}}

\newcommand{\calR}{\mathcal{R}}

\def\diag{\mathop{\rm diag}\nolimits}

\begin{document}

\begin{frontmatter}

\title{High-dimensional generalized semiparametric model for longitudinal data}

\author{M. Taavoni and M. Arashi\fnref{Corresponding author.}}
\address{Department of Statistics, Faculty of Mathematical Sciences, Shahrood University of Technology, Shahrood, IRAN}
\fntext[myfootnote]{Corresponding author.}




\begin{abstract}
This paper considers the problem of estimation in the generalized semiparametric model for longitudinal data when the number of parameters diverges with the sample size. A penalization type of generalized estimating equation method is proposed, while we use the regression spline to approximate the nonparametric component. The proposed procedure involves the specification of the posterior distribution of the random effects, which cannot be evaluated in a closed-form. However, it is possible to approximate this posterior distribution by producing random draws from the distribution using a Metropolis algorithm. Under some regularity conditions, the resulting estimators enjoy the oracle properties, under the high-dimensional regime. Simulation studies are carried out to assess the performance of our proposed method, and two real data sets are analyzed to illustrate the procedure.
\end{abstract}


\end{frontmatter}


\section{Introduction}\label{sec:intro}
Longitudinal studies are often conducted in epidemiology, social science and other biomedical research areas. A challenge in the analysis of longitudinal data is that the repeated measurements from the same subjects are correlated over time. A popular way for incorporating this correlation within the likelihood framework is to use the linear mixed-effects model (LMM; \cite{Laird:Ware:1982}) to analyze continuous longitudinal data and the generalized linear mixed-effects model (GLMM; \cite{Zeger:Karim:1991}) to analyze discrete longitudinal data, where the random component takes care the correlation among observations from the same subjects. However, the traditional GLMM assumes parametric fixed-effects that may be too restrictive to account complex covariate effects, especially when the variety of response over time is in a complicated manner.

To eliminate the limitation of the GLMMs for modeling non linear time trend, a generalized semiparametric mixed-effects model (GSMM), a natural extension of the GLMMs and semiparametric mixed models (SMM; \cite{Zeger:Diggle:1994}), is widely used to analyze longitudinal data by incorporating the within subject correlation using random effects and an arbitrary smooth function to model the time effect. Further developments along this line in the framework of GSMM can be found in \cite{Fan:et:al:2007,Qin:Zhu:2007,Qin:Zhu:2009,Kurum:et:al:2016} to mention a few.

There is a large body of variable selection methods for cross-sectional data. Among all, we refer to bridge regression \citep{Frank:Friedman:1993},  Lasso \citep{Tibshirani:1996},  adaptive Lasso \citep{Zou:2006}, Elastic-net \citep{Zou:Hastie:2005}, and SCAD \citep{Fan:Li:2001}. The literature on variable selection for longitudinal data is rather limited due to the challenges imposed by incorporating the intracluster correlation. \cite{Fan:Li:2004} extended the SCAD procedure to the semiparametric model for longitudinal data. \cite{Bondell:et:al:2010} proposed simultaneous selection of the fixed and random factors using a penalized joint log likelihood for the LMM. \cite{Ni:et:al:2010} proposed a double-penalized likelihood approach for simultaneous model selection and estimation for the SMM. \cite{Ma:et:al:2013} applied proper penalty functions in the additive semiparametric model. \cite{Chu:et:al:2016} developed a screening procedure for ultrahigh dimensional longitudinal data. In contrast to extensive attention on model selection for Gaussian longitudinal data, research on model selection for non-Gaussian longitudinal data in the framework of the GLM remains largely unexplored. To do variable selection, \cite{Pan:2001} developed a quasi-likelihood information criterion (QIC) which is analogous to AIC; \cite{Cantoni:et:al:2005} generalized Mallow's $C_p$ criterion, and \cite{Wang:Qu:2009} proposed a BIC criterion based on the quadratic inference function. These are best subset type model selection procedures which become computationally intensive when the number of parameters is moderately large.  Regarding regularization methods for longitudinal data, \cite{Fu:2003} proposed a generalization of the bridge and Lasso penalties to the generalized estimating equations (GEE) model. \cite{Xu:Zhu:2010} extended the independence screening method to deal with the high dimensional longitudinal GLMs. \cite{Dziak:2006} generalized the Lasso and SCAD methods to the longitudinal GLMs. The SCAD-penalized selection procedures were illustrated in \cite{Xue:et:al:2010} for the generalized additive model with correlated data. In all aforementioned studies on the penalized GLM for longitudinal data, the dimension of predictors is fixed. \cite{Xu:et:al:2012} proposed a weighted least squares type function to study the longitudinal GLMs with a diverging number of parameters. For correlated discrete outcome data, the joint likelihood function does not have a closed form if the correlation information is taken into account. When the dimension of parameters diverges, numerical approximation to the joint likelihood function tends to be computationally infeasible as it often involves high-dimensional integration. This motivated \cite{Liang: Zeger:1986} to develop an approach of the GEE which is a multivariate analogue of the quasi-likelihood. \cite{Johnson:et:al:2008} recently derived the asymptotic theory for the penalized estimating equations for independent data. \cite{Wang:et:al:2012} employed rather different techniques than those in \cite{Johnson:et:al:2008} and proposed the SCAD-penalized GEE for analyzing longitudinal data with high dimensional covariates. To the best of our knowledge, regularization in the GSMM is neglected.

In this paper, we focus on the GSMM with longitudinal data by allowing for non-Gaussian data and nonlinear link function. We consider the case where the number of variables $p$ is allowed to increase with the number of sample size $n$. Similar to the work of \cite{Wang:et:al:2012}, we apply the penalty function to the estimating equation objective function. Our method is rather different from their work because of including random effects and a nonparametric component in the model. We adopt spline regression to estimate the nonparametric components. The proposed penalized estimation involves the specification of the posterior distribution of the random effects, which cannot be evaluated in a closed form. However, it is possible to approximate this posterior distribution by producing random draws from a distribution using the Metropolis algorithm \citep{Tanner:1993}, which does not require the specification of the posterior distribution. We establish the asymptotic theory for the proposed method in a high-dimensional framework where the number of covariates increases with the sample size. To estimate the parameters, a computationally flexible iterative algorithm is developed. Furthermore, we propose a sandwich formula to estimate the asymptotic covariance matrix. 

The rest of this paper is organized as follows. Section \ref{sec2}, formulates the model and considers the estimation under the GEE framework. 
Section \ref{sec3} includes selection of the regularization parameters and the model selection procedure. Furthermore, asymptotic properties of the estimators are studied.
In Section \ref{sec4}, we apply a number of simulations to assess the finite sample performance of the proposed estimation method in the GSMM. A real data analysis is also presented in this section to augment the theoretical results. Some concluding remarks are given in Section \ref{sec5}. 
Further, the proofs of the main results as well as some instrumental lemmas are provided in a separate supplementary file.

\section{Generalized Semiparametric Model}\label{sec2}
\subsection{Model specification}
Consider a longitudinal study with $n$ subjects and $n_i$ observations over time for the $i$th subject ($i = 1, \ldots, n$).
Let $\bu_i$ be a $q\times 1$ vector of random effects corresponding to the $i$th subject, and $y_{ij}$ be an observation of the $i$th subject measured at time $t_{ij}$ for $i = 1,\ldots, n$ and $j = 1,\ldots, n_i$. Suppose that $y_{i1}, \ldots, y_{in_i}$ given $\bu_i$ are conditionally independent and each $y_{ij}|\bu_i$ is distributed as an exponential family distribution whose probability density function is given by
\begin{eqnarray}\label{eq1}
p(y_{ij}|\bu_i, \bbeta_n, \phi)=\exp\left[\phi^{-1}\lbrace y_{ij}\theta_{ij}- b(\theta_{ij})\rbrace +c(y_{ij}, \phi)\right],
\end{eqnarray}
where $\phi$ is a scale parameter, $c(., .)$ is a function only depending on $y_{ij}$ and $\phi$, and $\theta_{ij}$ is the (scalar) canonical parameter. The conditional expectations and variances of $y_{ij}$ given $\bu_{i}$ are given by $\mu_{ij}=E(y_{ij}|\bu_{i})=b^{.}(\theta_{ij})$ and $\nu_{ij}= var(y_{ij}|\bu_{i})=\phi b^{..}(\theta_{ij})$, respectively, where $b^{.}(\theta)=\frac{\partial b(\theta)}{\partial \theta}$ and $b^{..}(\theta)=\frac{\partial^2 b(\theta)}{\partial \theta^2}$. In this paper, we assume that the conditional mean $\mu_{ij}$ satisfies
\begin{eqnarray}\label{eq2}
g(\mu_{ij})\triangleq \eta_{ij}= \bX^\top_{ij}\bbeta_n +\bZ^\top_{ij}\bu_i+ f(t_{ij}), \quad i = 1,\ldots, n;  ~j = 1,\ldots, n_i,
\end{eqnarray}
where $g(.)$ is a known monotonic link function, $\bX^\top_{ij}$ is a $p_n\times 1$ vector of explanatory variables, $\bbeta_n$ is a $p_n\times 1$ vector of unknown parameters of the fixed effects, $\bZ^\top_{ij}$ is a $q\times 1$ vector relating to the random effects and treated as a subset of fixed effects excluding time variables, $f(.)$ is an unknown smooth function which is continuous and twice differentiable function on some finite interval. The dimension of the covariates
$p_n$ is allowed to depend on the number of subjects $n$. To complete the specification, assume that the random effects $\bu = \lbrace\bu_1, \ldots , \bu_q\rbrace$ independently follow the same distribution, depending on parameters $\bSigma$ as
\begin{eqnarray}\label{eq3}
\bu_i\thicksim f_u(\bu_i|\bSigma).
\end{eqnarray}
The model defined in Eqs. \eqref{eq1}--\eqref{eq3} is referred to as generalized semiparametric mixed model (GSMM). Specific assumptions will be considered for the number of variables $p_n$ in section \ref{sec asymp}.

We approximate the unspecified smooth function using
\begin{eqnarray*}
	f(t_{ij})=\alpha_{0}+\alpha_{1}t_{ij}+\ldots+\alpha_{d}t_{ij}^{d}+\sum_{l=1}^{L_n}\alpha_{(d+1)+l}(t_{ij}-t_{i}^{(l)})_{+}^{d}=\bB(t_{ij})^\top\balpha_n,
\end{eqnarray*}
where $d$ is the degree of the polynomial component, $L_n$ is the number of interior knots (rate of $L_n$ will be specified in Section \ref{sec asymp}), $t_{i}^{(l)}$ is  referred as knots of the $i$th subject, $\bB(t_{ij})=\Big( 1, t_{ij},\ldots, t_{ij}^{d}, \big(t_{ij}-t_{i}^{(1)}\big)^{d}_+ , \ldots , \big(t_{ij}-t_{i}^{(L_n)}\big)^{d}_+\Big )$ is a $h_n \times 1$ vector of basis functions, $h_n$ is the number of basis functions used to approximate $f(t_{ij})$, $h_n=d+1+L_n$ , $(a)_{+} = \hbox{max}(0, a)$, and $\balpha_n=(\alpha_{0}, \ldots,\alpha_{d},\alpha_{d+1}, \ldots,\alpha_{d+L_n})^\top$ is the spline coefficients vector of dimension $h$.  Thus, we can represent the regression model \eqref{eq2} as
\begin{eqnarray}\label{eq4}
\eta_{ij}= \bX^\top_{ij}\bbeta_n +\bZ^\top_{ij}\bu_i+ \bB(t_{ij})^\top\balpha_n, \quad i = 1,\ldots, n;  j = 1,\ldots, n_i.
\end{eqnarray}
For convenience, model \eqref{eq4} can take the form
$\eta_{ij}= \bD^\top_{ij}\btheta_n +\bZ^\top_{ij}\bu_i$,
where $\bD_{ij}=\big(\bX_{ij}^\top,\bB_j(\bt_i)^\top\big)^\top$ being a $(p_n+h_n)\times 1$ design matrix combining the fixed-effects and spline-effects design matrices for the $j$th outcome of the $i$th subject, and $\btheta_n=(\bbeta_n^\top, \balpha_n^\top)^\top$ is a $(p_n+h_n)\times 1$ combined regression parameters vector that must be estimated.

Now linearization of the GSMM can be formulated in the seamless form
\begin{eqnarray}\label{eq5}
p(y_{ij}|\bu_i, \btheta_n, \phi)&=&\exp\left[\phi^{-1}\lbrace y_{ij}\theta_{ij}- b(\theta_{ij})\rbrace +c(y_{ij}, \phi)\right],\cr
\bu_i&\thicksim &f_u(\bu_i|\bSigma),\quad 
\mu_{ij}=\mathbb{E}(y_{ij}|\bu_{i}),\cr
\eta_{ij}&=& \bD^\top_{ij}\btheta_n +\bZ^\top_{ij}\bu_i, \quad i = 1,\ldots, n;  j = 1,\ldots, n_i.
\end{eqnarray}

\subsection{Estimation procedure}

For linearization of the GSMM defined in \eqref{eq5}, the classical likelihood function can be defined as
\begin{eqnarray}\label{eq6}
L(\btheta_n, \bSigma, \phi)=\prod_{i=1}^n\int p_{\by_{i}|\bu_i}(\by_{i}|\bu_i, \btheta_n, \phi)p_{\bu}(\bu_i|\bSigma)d\bu_i
\end{eqnarray}
where $\by_{i}=(\by_{i1}, \ldots,\by_{in_i})^\top$, $\bu=(\bu_1, \ldots, \bu_n)  $, and
$
p_{\by_{i}|\bu_i}(\by_{i}|\bu_i, \btheta_n, \phi)=\prod_{j=1}^{n_i} p(\by_{ij}|\bu_i, \btheta_n, \phi)$.
For the maximum likelihood (ML) estimate we set up an EM algorithm and consider the random effects, $\bu_i$, to be the missing data. The complete data, is then ($\by_{i},\bu_i$) and the complete data log-likelihood is given by
\begin{eqnarray}\label{eq7}
\ell(\btheta_n, \bSigma, \phi)=\sum_{i=1}^n \hbox{ln} p_{\by_{i}|\bu_i}(\by_{i}|\bu_i, \btheta_n, \phi)+\sum_{i=1}^n \hbox{ln} p_{\bu_i}(\bu_i|\bSigma).
\end{eqnarray}
Using the separation as in \eqref{eq7}, the ML equations for $\btheta_n$ and $\bSigma$ take the forms
$\mathbb{E}\Big[\frac{\partial \hbox{ln}p_{y_{ij}|\bu_i}(y_{ij}|\bu_i,\btheta_n)}{\partial \btheta_n}|y_{ij}\Big]=\0$ and $\mathbb{E}\Big[\frac{\partial \hbox{ln}p_{\bu_i}(\bu_{i}|\bSigma)}{\partial \bSigma}|y_{ij}\Big]=\0.$
Using the Monte Carlo Newton-Raphson (MCNR) algorithm of \cite{McCulloch:1997}, the optimal estimating equation for $\btheta_n$ is given by
\begin{eqnarray}\label{eq8}
\mathbb{E}_{\bu|\by} \Big[n^{-1}\sum_{i=1}^n\frac{\partial \bmu_i(\btheta_n, \bu_i)}{\partial \btheta_n^\top}\bV_i^{-1}(\btheta_n, \bu_i)\big(\by_i-\bmu_i(\btheta_n , \bu_i)\big)\Big]=\0,
\end{eqnarray}
where $\bmu_i(\btheta_n, \bu_i)=(\mu_{i1},\ldots,\mu_{in_i})^\top$ and $\bV_i(\btheta_n, \bu_i)$ is the covariance matrix of $\by_{i}|\bu_i$. In real applications the true intracluster covariance structure is often unknown. The GEE procedure adopts a working covariance matrix, which is specified through a working correlation matrix 
$\bR(\brho): \bV_i(\btheta_n, \bu_i) = \bA_i^{\frac{1}{2}}(\btheta_n , \bu_i )\bR(\brho)\bA_i^{\frac{1}{2}}(\btheta_n , \bu_i )$, where $\brho$ is a finite dimensional parameter and $\bA_i(\btheta_n , \bu_i )=\diag(\nu_{i1},\ldots,  \nu_{in_i})$. With the estimated working correlation matrix $\widehat{\bR}\equiv \bR(\widehat{\brho}) $, the estimating equations in \eqref{eq8} reduces to
\begin{eqnarray}\label{eq9}
\mathbb{E}_{\bu|\by} \Big[n^{-1}\sum_{i=1}^n \bD_i^\top  \bA_i^{\frac{1}{2}}(\btheta_n , \bu_i ) \widehat{\bR}^{-1} \bA_i^{-\frac{1}{2}}(\btheta_n , \bu_i )\big(\by_i-\bmu_i(\btheta_n , \bu_i)\big)\Big]=\0,
\end{eqnarray}
where $\bD_i=(\bD_{i1}^\top, \ldots, \bD_{in_i}^\top)^\top$. We formally define the estimator as the solution $\widehat{\btheta}_n$ of the above estimating equations. For ease of exposition, we assume $\phi=1$   and $n_i = m <\infty$ in the rest of the article. Extension of the methodology to the cases of unequal $n_i$ is straightforward. We vary the dimension of $\bA_i$ and replace $\widehat{\bR}$ by $\widehat{\bR}_i$, which is the $n_i \times n_i$ matrix using the specified working correlation structure and the corresponding
initial parameter $\brho$ estimator.

\section{Regularization in the GPLMM}\label{sec3}
In order to select important covariates and estimate them simultaneously, the log likelihood \eqref{eq8} is expanded to include the penalty term $\sum_{k=1}^{p_n} p_{\lambda_n}(|\beta_{nk}|)$ which yields the following penalized log likelihood
\begin{eqnarray}\label{eq10}
\ell^p(\bbeta_n, \balpha_n, \bD, \phi)=\sum_{i=1}^n \hbox{ln} p_{\by_{i}|\bu_i}(\by_{i}|\bu_i, \btheta_n)+\sum_{i=1}^n p_{\bu_i}(\bu_i|\bSigma)-n\sum_{k=1}^{p_n}p_{\lambda_n}(|\beta_{nk}|),
\end{eqnarray}
where $p_\lambda(|\beta_{nk}|)$ is any penalty function and $\lambda_n$ is a tuning parameter. Since the coefficients $\btheta_n$ depends to the first and third terms of \eqref{eq10}, we propose the penalized estimating equation
$
\bU_n(\btheta_n)=\bS_n(\btheta_n)-q_{\lambda_n}(|\bbeta_n|)^\top\hbox{sign}(\bbeta_n)$,
where
$\bS_n(\btheta_n)=\mathbb{E}_{u|y} \Big[\sum_{i=1}^n \bD_i^\top  \bA_i^{\frac{1}{2}}(\btheta_n , \bu_i ) \widehat{\bR}^{-1} \bA_i^{-\frac{1}{2}}(\btheta_n , \bu_i )\big(\by_i-\bmu_i(\btheta_n , \bu_i)\big)\Big]$, with $q_{\lambda_n}(|\bbeta_n|)=\big(q_{\lambda_n}(|\beta_{n1}|),\ldots,q_{\lambda_n}(|\beta_{np_n}|)\big)$ is a $1\times p_n $ vector of penalty functions, $\hbox{sign}(\bbeta_n) = \big(\hbox{sign}(\beta_{n1}), \ldots ,
\hbox{sign}(\beta_{np_n} )\big)$ with $\hbox{sign}(a) = I(a> 0) - I(a < 0)$ and $q_{\lambda_n}(|\beta_{nk}|) =p_{\lambda_n}^{'}(|\beta_{nk}|)$.

Note that we assume the semiparametric part contains significant contribution in the model and the proposed penalized estimating equation has been defined to shrink small components of the coefficient $\bbeta_n$ to zero not $\balpha_n$. Thus, the method performing variable selection for fixed effects, produces estimators of the nonzero components and the nonparametric component.

We use the SCAD penalty proposed by \cite{Fan:Li:2001} defined by
\begin{eqnarray*}
	q_{\lambda_n}(|\beta_n|) =p^{'}_{\lambda_n}(|\beta_n|)=\lambda_n\Big\lbrace  I(|\beta_n|\leq \lambda_n)+\frac{(a\lambda_n-|\beta_n|)_+}{(a-1)\lambda_n}I(|\beta_n|> \lambda_n) \Big\rbrace; \quad a>2,
\end{eqnarray*}
where the notation $(.)_+$ stands for the positive part of $(.)$. 

Our proposed estimator for $\btheta_n$ is the solution of $\bU_n(\btheta_n)=\0$. Because $\bU_n(\btheta_n)$ has discontinuous points, an exact solution to $\bU_n(\btheta_n)=\0$ may not exist. We formally define the estimator $\widehat{\btheta}_n$ to be an approximate solution, i.e., $\bU_n(\widehat{\btheta}_n)= o(a_n )$ for a sequence $a_n \to 0$. 
Alternatively, since the penalty function is singular at the origin, it is challenging to obtain the estimator of $\btheta_n$ by solving $U_n(\btheta_n)=\0$. In the neighborhoods of the true parameter values $\beta_{n0k}$, $|\beta_{n0k}|>0$, the derivative of the penalty function is well approximated by
\begin{eqnarray*}
	q_{\lambda_n}(|\beta_{nk}|)\hbox{sign}(\beta_{nk})\thickapprox \frac{q_{\lambda_n}(|\beta_{n0k}|)}{|\beta_{n0k}|}\beta_{nk}.
\end{eqnarray*}
With the local quadratic approximation, we apply the Newton-Raphson method to solve $\bU_n(\widehat{\btheta}_n)= o(a_n )$, and get the following updating formula
\begin{eqnarray}\label{eq12}
\widehat{\btheta}_n^{(m+1)}=\widehat{\btheta}_n^{(m)}+
\Big \lbrace \bH_n(\widehat{\btheta}_n^{(m)})+n \bE_n(\widehat{\btheta}_n^{(m)})\Big\rbrace^{-1} \times
\Big\lbrace  \bS_n(\widehat{\btheta}_n^{(m)})+n \bE_n(\widehat{\btheta}_n^{(m)})\widehat{\btheta}_n^{(m)} \Big\rbrace,
\end{eqnarray}
where 
\begin{eqnarray*}
	\bH_n(\widehat{\btheta}_n^{(m)})=\mathbb{E}_{u|y} \Big[\sum_{i=1}^n \bD_i^\top  \bA_i^{\frac{1}{2}}(\btheta_n , \bu_i ) \widehat{\bR}^{-1} \bA_i^{\frac{1}{2}}(\btheta_n , \bu_i )\bD_i\Big],
\end{eqnarray*}
\begin{eqnarray*}
	\bE_n(\widehat{\btheta}_n^{(m)})=\hbox{diag}\Big \lbrace \frac{q_{\lambda_n}(|\beta_{n1}|)}{\epsilon+|\beta_{n1}|},  \ldots, \frac{q_{\lambda_n}(|\beta_{np_n}|)}{\epsilon+|\beta_{np_n}|},\0_{h_n}\Big \rbrace,
\end{eqnarray*}
for a small numbers e.g. $\epsilon=10^{-6}$. Here, and $\0_{h_n}$ denotes a zero vector of dimension $h_n$.

In the forthcoming section we outline the computational procedure used for sample generation. 

\subsection{MCNR algorithm}
Let $\bU$ denote the previous draw from the conditional distribution of $\bU|\by$, and generate a new value $u_k^*$ for the $j$th component of $\bU^*=(u_1, \ldots , u_{k-1}, u^*_k, u_{k+1}, \ldots, u_{nq})$ by using the candidate distribution $p_{\bu}$, accept $\bU^*$ as the new value with probability
\begin{eqnarray}\label{eq13}
\alpha_k(\bU,\bU_*)=\hbox{min}\Big\lbrace 1,  \frac{p_{u|y}(\bU^*|\by,\btheta_n,\bD)p_{u}(\bU|\bD)}{p_{u|y}(\bU|\by,\btheta_n,\bD)p_{u}(\bU^*|\bD)} \Big\rbrace.
\end{eqnarray}
otherwise, reject it and retain the previous value $\bU$. The second term in brace in \eqref{eq12} can be simplified to
\begin{eqnarray*}
	\frac{p_{\bu|\by}(\bU^*|\by,\btheta_n,\bD)p_{u}(\bU|\bD)}{p_{\bu|\by}(\bU|\by,\btheta_n,\bD)p_{u}(\bU^*|\bD)}&=&\frac{p_{\by|\bu}(\by|\bU^*,\btheta_n)}{f_{\by|\bu}(\by|\bU,\btheta_n)}\\ \nonumber
	&=&\frac{\prod_{i=1}^n p_{\by_i|\bu}(\by_i|\bU^*,\btheta_n)}{\prod_{i=1}^n f_{\by_i|\bu}(\by_i|\bU,\btheta_n)}.
\end{eqnarray*}
Note that, the calculation of the acceptance function $\alpha_k(\bU,\bU_*)$ here involves only the specification of the conditional distribution of $\by|\bu$ which can be computed in a closed form.

\subsection{Choice of regularization parameters}
For computational convenience, we use equally spaced knots with the number of interior knots $L_n\approx  n^{1/(2r+1)}$, where $r$ is positive integer. A similar strategy for knot selection can also be found in \cite{He:et:al:2002, Qin:Zhu:2007,Sinha:Sattar:2015}.
To reduce the computational burden, we follow \cite{Fan:Li:2001} and set $a=3.7$. 
To select the tuning parameter $\lambda_n$ we use the GCV suggested by \cite{Fan:Li:2001} given by
\begin{eqnarray*}
	\hbox{GCV}_{\lambda_n}=\frac{\hbox{RSS}(\lambda_n)/n}{(1-d(\lambda_n)/n)^2},
\end{eqnarray*}
where 
\begin{eqnarray}\label{eq14}
\hbox{RSS}(\lambda_n)=\frac{1}{N}\sum_{k=1}^N\Big[  \sum_{i=1}^n \big(\by_i-\bmu_i(\widehat{\btheta}_n , U_i^{(k)})\big)^\top\bW_i^{-1}\big(\by_i-\bmu_i(\widehat{\btheta}_n , U_i^{(k)})\big) \Big]\nonumber\\
\end{eqnarray} 
is the residual sum of squares, and 
\begin{eqnarray*}
	d(\lambda_n)=tr\Big[ \Big \lbrace \frac{1}{N}\sum_{k=1}^N\Big[ \bH_n\big(\widehat{\btheta}_n, \bU^{(k)}\big)\Big]+n \bE_n(\widehat{\btheta}_n)\Big\rbrace^{-1}\times \Big \lbrace \frac{1}{N}\sum_{k=1}^N\Big[ \bH_n\big(\widehat{\btheta}_n, \bU^{(k)}\big)\Big]\Big\rbrace \Big]
\end{eqnarray*} 
is the effective number of parameters. Then, $\lambda_{opt}$ is the minimizer of the $\hbox{GCV}_{\lambda_n}$.
Note that $\bW_i$ in \eqref{eq14} is an $n_i\times n_i$ covariance matrix of $\by_i$, that can be computed as
$
\bW_i=\mathbb{E}_{u|y}  \Big(\hbox{var}(\by_i|\bu_i)\Big)+\hbox{var}_{u|y}  \Big(\mathbb{E}(\by_i|\bu_i)\Big),
$
where
\begin{eqnarray*}
	\mathbb{E}_{u|y}  \Big(\hbox{var}(\by_i|\bu_i)\Big)&=&\frac{1}{N}\sum_{k=1}^N\Big[ \bV_i(\widehat{\btheta}_n, U_i^{(k)}) \Big],\\
	\hbox{var}_{u|y}  \Big(\mathbb{E}(\by_i|\bu_i)\Big)&=&\frac{1}{N}\sum_{k=1}^N\Big[\bmu_i(\widehat{\btheta}_n , U_i^{(k)})\Big]^2-\bigg[ \frac{1}{N}\sum_{k=1}^N\Big[\bmu_i(\widehat{\btheta}_n , U_i^{(k)})\Big]\bigg]^2.
\end{eqnarray*}

\subsection{Asymptotic properties}\label{sec asymp}
Assume the true value of $\bbeta_{0}$ is partitioned $\bbeta_{0}=(\bbeta_{01}^\top,\bbeta_{02}^\top)^\top$ and the corresponding design matrix into $\bX_i = \big(\bX_{i(1)}, \bX_{i(2)}\big)$. In our study, the true regression coefficients are $\btheta_{n0}=(\bbeta_{01}^\top,\bbeta_{02}^\top,\balpha_{0}^\top)^\top$, where $\balpha_{0}$ is an $h_n$-dimensional vector depending on $f_0$. For technical convenience let $\btheta_{0}=(\btheta_{01}^\top,\btheta_{02}^\top)^\top$ where $\btheta_{01}=(\bbeta_{01}^\top,\balpha_{0}^\top)^\top$ is  $(s=s^*+h_n)$-dimensional vector of true values that the elements are all nonzero and $\btheta_{02}=\bbeta_{02}=\0$.
Here, $s^*$ is the dimension of $\btheta_{01}$ and assume that only a small number of covariates contribute to the response i.e. $\mathcal{S}=\lbrace 1\leq j \leq p ; \beta_j \neq 0 \rbrace$ has cardinal $ |\mathcal{S}|=s^*<p $. 
Consequently, estimated values and the design matrix is repartitioned as $\widehat{\btheta}_n=(\widehat{\btheta}_{n1}^\top,\widehat{\btheta}_{n2}^\top)^\top$, and $\bD_{i}=\big(\bD_{i(1)}^\top, \bD_{i(2)}^\top\big)^\top$ which $\widehat{\btheta}_{n1}=(\widehat{\bbeta}_{n1}^\top,\widehat{\balpha}_n^\top)^\top$, $\bD_{i(1)}=\big(\bX_{i(1)}^\top, \bB(\bt_i)^\top\big)^\top$, $\widehat{\btheta}_{n2}=\widehat{\bbeta}_{n2}$  and $\bD_{i(2)}=\bX_{i(2)}$.

Meanwhile, If Eq. \eqref{eq9} has multiple solutions, only a sequence of consistent estimator $\widehat{\btheta}_n$ is considered. 

Throughout, we need some regularity conditions and to save space we managed to put them in a separate supplementary file (SF) and refer here as (A.1)-(A.8). Further, some lemmas are also used that we put them in the SF. 
Now, consider the following estimating equation
\begin{eqnarray*}
	\overline{\bS}_n(\btheta)=\mathbb{E}_{u|y} \Big[\sum_{i=1}^n \bD_i^\top  \bA_i^{\frac{1}{2}}(\btheta , \bu_i ) \overline{\bR}^{-1} \bA_i^{-\frac{1}{2}}(\btheta , \bu_i )\big(\by_i-\bmu_i(\btheta , \bu_i)\big)\Big].
\end{eqnarray*}
Let $\overline{\bM}_n(\btheta_n)$ to be the covariance matrix of $\overline{\bS}_n(\btheta)$, then
\begin{eqnarray*}
	\overline{\bM}_n(\btheta)=\mathbb{E}_{u|y} \Big[\sum_{i=1}^n \bD_i^\top  \bA_i^{\frac{1}{2}}(\btheta , \bu_i ) \overline{\bR}^{-1}\bR_0 \overline{\bR}^{-1} \bA_i^{\frac{1}{2}}(\btheta , \bu_i )\bD_i\Big].
\end{eqnarray*}
By Lemma 1 (see the SF), we approximate $f_0(t)$ by $\bB(t)\balpha_0$, then have
\begin{eqnarray*}
	\eta_{ij}(\btheta_0)=g\big(\mu_{ij}(\btheta_0)\big)=\bX_{ij}^\top\bbeta_0+\bB(t_{ij})\balpha_0+\bZ_{ij}^\top\bu_i, \quad \btheta_0=(\bbeta_0^\top,\balpha_0^\top)^\top_{(p_n+N)\times 1}.
\end{eqnarray*}
Theorems \ref{th1}-\ref{th3} below characterize the existency, consistency and normality of the proposed penalized estimator when $p_n\to\infty$.

\begin{theorem}\label{th1}
	(Existency). Assume the conditions (A.1)--(A.8). Then, there exists an approximate penalized GEE solution $\widehat{\btheta}=(\widehat{\btheta}_{1}^\top,\widehat{\btheta}_{2}^\top)^\top$
	which satisfies the following properties
	\begin{eqnarray*}
		\textrm{(i)}&& \mathbb{P}_n \Big(\vert U_{nk}(\widehat{\btheta}_n) \vert=0 , ~k=1, \ldots, s_n^*, (s_n^*+1),\ldots, (s_n=s_n^*+h_n)\Big)\to 1,\cr
		\textrm{(ii)}&& \mathbb{P}_n\Big(\vert U_{nk}(\widehat{\btheta}_n) \vert \leq \frac{\lambda_n}{\log n} , ~k=(s_n^*+h_n+1), \ldots, p_n\Big)\to 1,
	\end{eqnarray*}
	where
	\begin{eqnarray*}
		U_{nk}(\widehat{\btheta}_n)=\left\lbrace
		\begin{array}{cc}
			S_{nk}(\widehat{\btheta}_n)-n\frac{q_{\lambda_n}(|\widehat{\beta}_{nk}|)}{\epsilon+|\widehat{\beta}_{nk}|}\widehat{\beta}_{nk}& \quad k=1,\ldots,s_n,\\
			S_{nk}(\widehat{\btheta}_n)& \quad k=(s_n+1),\ldots, p_n\\
		\end{array},\right.
	\end{eqnarray*}
	and $S_{nk}(\widehat{\btheta}_n)$ denotes the $k$th element of $\bS_{n}(\widehat{\btheta}_n)$.
\end{theorem}
\begin{theorem}\label{th2}
	(Consistency). Assume conditions (A1)--(A8) and that $n^{-1}p_n^2= o(1)$. Then, $\bU_n(\btheta_n)= o(1)$ has a root $\widehat{\btheta}_n$ such that
	\begin{eqnarray*}
		(i)&&\Vert \widehat{\btheta}_n-\btheta_{n0} \Vert = O_p(\sqrt{p_n/n}),\cr
		(ii)&& \frac{1}{n}\sum_{i=1}^n\sum_{j=1}^{n_i}\big(\widehat{f}(t_{ij})-f_0(t_{ij})\big)^2=O_p(n^{-2r/(2r+1)}).
	\end{eqnarray*}
\end{theorem}
\begin{theorem}\label{th3}
	(Oracle properties). Assume (A.1)--(A.8). If $L_n \approx n^{1/(2r+1)}$, and $n^{-1}p_n^3 = o(1)$, then $\forall \bxi_n \in \calR^{p_n}$ such that $\Vert \bxi_n \Vert=1$,  we have
	\begin{eqnarray*}
		(i)&& \mathbb{P}_n(\widehat{\bbeta}_{n2}=\0)\to 1,\cr
		(ii)&& \bxi_n^\top\overline{\bM}_n^{*^{-1/2}}(\bbeta_{n0})\overline{\bH}_n^*(\bbeta_{n0})(\widehat{\bbeta}_{n1}-\bbeta_{n01})\overset{\mathcal{D}}{\to} \hbox{N}_{p_n}(0,1),
	\end{eqnarray*}
	where
	\begin{eqnarray*}
		\overline{\bM}_n^{*}&=&\mathbb{E}_{u|y} \Big[\sum_{i=1}^n \bX_i^{*^\top}  \bA_i^{\frac{1}{2}}(\btheta_n , \bu_i ) \overline{\bR}^{-1}\bR_0 \overline{\bR}^{-1} \bA_i^{\frac{1}{2}}(\btheta_n , \bu_i )\bX_i^{*}\Big],\\
		\overline{\bH}_n^{*}&=&\mathbb{E}_{u|y} \Big[\sum_{i=1}^n \bX_i^{*^\top}  \bA_i^{\frac{1}{2}}(\btheta_n , \bu_i ) \overline{\bR}^{-1} \bA_i^{\frac{1}{2}}(\btheta_n , \bu_i )\bX_i^{*}\Big],
	\end{eqnarray*}
	$\bX_i^{*}=(\bI-\bP)\bX_i$, $\bP=\bB(\bB^\top\bOmega\bB)^{-1}\bB^\top\bOmega$, $\bOmega=\hbox{diag}\lbrace \bOmega_i\rbrace$ and $\bOmega_i=\mathbb{E}_{u|y} \Big[\bA_i^{\frac{1}{2}}(\btheta_n , \bu_i ) \overline{\bR}^{-1} \bA_i^{\frac{1}{2}}(\btheta_n , \bu_i )\Big]$.
\end{theorem}

To estimate the the asymptotic covariance matrix of $\widehat{\btheta}_n$, we use the following sandwich formula:
\begin{eqnarray*}
	\hbox{Cov}(\widehat{\btheta}_n)\approx[\bH_n(\widehat{\btheta}_n,\bu_i)
	+n\bE_n(\widehat{\btheta}_n)]^{-1} \bM_n(\widehat{\btheta}_n,\bu_i) [\bH_n(\widehat{\btheta}_n,\bu_i)+n\bE_n(\widehat{\btheta}_n,\bu_i)]^{-1},
\end{eqnarray*}
where $\bH_n$ and $\bE_n$ are defined in Section \ref{sec3}, and
\begin{eqnarray*}
	\bM_n(\widehat{\btheta}_n,\bu_i)= \sum_{i=1}^n\bD_i^\top\bA_i^{1/2}(\widehat{\btheta}_n,\bu_i)\widehat{\bR}^{-1} \big[\bepsilon_i(\widehat{\btheta}_n,\bu_i)\bepsilon_i^\top(\widehat{\btheta}_n,\bu_i)\big]  \widehat{\bR}^{-1}\bA_i^{1/2}(\widehat{\btheta}_n,\bu_i)\bD_i^\top.
\end{eqnarray*}

\section{Numerical Studies}\label{sec4}
\subsection{Simulation}
We generated 100 data sets following $
y_{ij}|b_i \thicksim \hbox{Pois}(\mu_{ij})$, with $\eta_{ij}=\hbox{log}(\mu_{ij})=\sum_{k=1}^p x_{ij}^{(k)}\beta_k+\hbox{sin}(2\pi t_{ij})+b_i$,  
where $i = 1, \ldots , n$ ($n = 50, 100$ and 150), and $j = 1, \ldots, n_i$ which the number of observations per subjects assumed to be fixed at $n_i=5$. The true regression coefficients are $\bbeta=(-1,-1,2, 0,\ldots,0)$ with the mutually independent covariates $\bX_{ij}^\top=(x_{ij}^{(1)},\ldots, x_{ij}^{(p)})$ are drawn independently from uniform distribution on $(-1,1)$. The measurement time points $t_{ij}$ are drawn from uniform distribution on $(0,1)$. The random effect process $b_i$ is taken to be a Gaussian process with mean 0, variance $\sigma^2=0.25$. The predictor dimension $p_n$ is diverging but the dimension of the true model is fixed to be 3.

Regarding the choice of the dimensionality of the parametric component, $p_n$, authors recommended many suggestions as a sensible choice. For example $p_n=[\frac{n}{2}]$, $p_n = [4.5n^{1/4}]$, and $p_n = [\frac{n}{b\log(n)}]$, where $b>1$ and $[a]$ stands for the largest integer no larger than $a$. These only discuss the situation $p\to \infty$ as $n\to \infty$ with $p_n < n$. For case $p_n>>n$, we can mention to $\log(p_n) = o_p(n^b)$, where $0<b<1$. Of course challenges arise when $p$ is much larger than $n$, choosing a larger value of $p_n$ increases the probability that variable selection methods will include all of the correct variables, but including more inactive variables will tend to have a slight detrimental effect on the performance of the final variable selection and parameter estimation method. We have found that this latter effect is most noticeable in models where the response provides less information. We therefore used the pairs of $(n,p_n)$ as $(50,11), (100,14), (150,16)$ and $(30,100), (100, 500), (200,2000)$ respectively for cases $p_n < n$ and $p_n>>n$. 

Performance of the proposed penalized procedure compared with the unpenalized one and the penalized GLMM where each simulated data set was fitted under these three methods. For evaluating estimation accuracy, we report the empirical mean square error (MSE), defined as $\sum_{k=1}^{100}\Vert\widehat{\bbeta}_n^k-\bbeta_{n0}\Vert/100$ where $\widehat{\bbeta}_n^k$ is the estimator of $\bbeta_{n0}$ obtained using the $k$th generated data set. The performance in variable selection is gauged by (C, I), where `C' is the mean over all 100 simulations of zero coefficients which are correctly estimated by zero and `I' is the mean over all 100 simulations of nonzero coefficients which are incorrectly estimated by zero. To present a more comprehensive picture, we also use other criteria for variable selection performance evaluation. `Under-fit' corresponds to the proportion of excluding any true nonzero coefficients. Similarly, we report the proportion of selecting the exact subset model as `Correct-fit' and the proportion of including all three important variables plus some noise variables as `Over-fit'.  

The results of Table \ref{tab1} summarize the estimation accuracy and model selection properties of the penalized GSMM (P-GSMM), the unpenalized GSMM, and the penalized GLMM (P-GLMM) for the different values of $(n,p_n)$. In terms of estimation accuracy the penalized GSMM procedure performs closely to the penalized GLMM, whereas our proposed approach gives the smallest MSE, and consistently outperforms its penalized GLMM counterpart. In terms of model selection we observe that the unpenalized GSMM generally does not lead to a sparse model. Furthermore, the penalized GSMM and the penalized GLMM successfully selects all covariates with nonzero coefficients (i.e., I rates are zero), but it is obvious that the proposed approach has slightly stronger sparsity (i.e., a fairly higher number of Cs) than the penalized GLMM. For penalized GSMM, The probability of identifying the exact underlying model is about $80\%$ and this rate grows by increasing the sample size, confirming the good asymptotic properties of the penalized estimators. The results are the same in both cases of $p_n<n$ and $p_n>>n$, but when $p_n>>n$ zero coefficients tends to increasingly included in the model.

To further investigate the performance of the proposed method, Table \ref{tab2} reports its bias, the estimated standard deviation (calculated from the sandwich variance formula), the empirical standard deviation, and the empirical coverage probability of $95\%$ confidence interval for estimating $\beta_1$, $\beta_2$, and $\beta_3$. The estimated standard deviation is close to the empirical standard deviation, and the empirical coverage probability is close to $95\%$. This indicated good performance of the sandwich variance formula. 

These observations suggest that considering partial part is important to modify the estimation accuracy and model selection when the growth curves of the data exhibit a nonlinear fashion over time, especially in a complicated manner. On the other hand, penalized GSMM allows us to make systematic inference on all model parameters by representing a partially model as a modified penalized GLMM.

\begin{table}[h!]
			\small
			 \begin{adjustwidth}{-2cm}{}
			\caption{Model selection results for Poisson responses: comparison of P-GSMM, GSMM, and P-GLMM with the cases of $p_n<n$ and $p_n>>n$}\label{tab1}
			\begin{tabular}{ccccccccccccc}\hline
				\noalign{\hrule height 1pt}\\
				method&\multicolumn{6}{c}{case $p_n<n$}&\multicolumn{6}{c}{case $p_n>>n$}\\
				\cmidrule(r){2-7}\cmidrule(r){8-13}
				&\multicolumn{6}{c}{$(n,p)=(50,11)$}&\multicolumn{6}{c}{$(n,p)=(30,100)$}\\
				&MSE& C(8)&I(0)&Under-fit&Correct-fit&Over-fit
				&MSE& C(97)&I(0)&Under-fit&Correct-fit&Over-fit\\
				\cmidrule(r){2-7}\cmidrule(r){8-13}
				GPLMM&0.116&0.09&0.00&0.00&0.00&1.00&68.028&0.074&0.00&0.00&0.00&1.00 \\
				P-GLMM&0.060&6.54&0.00&0.00&0.13&0.87&0.435&96.48&0.00&0.00&0.55&0.45\\
				P-GPLMM&0.052&7.59&0.00&0.00&0.64&0.36&0.391&96.41&0.00&0.00&0.60&0.40\\
				\\
				&\multicolumn{6}{c}{$(n,p)=(100,14)$}&\multicolumn{6}{c}{$(n,p)=(100,500)$}\\
				&MSE& C(11)&I(0)&Under-fit&Correct-fit&Over-fit
				&MSE& C(497)&I(0)&Under-fit&Correct-fit&Over-fit\\
				\cmidrule(r){2-7}\cmidrule(r){8-13}
				GPLMM&0.072&0.16&0.00&0.00&0.00&1.00&1499.136&47.02&0.03&0.03&0.00&1.00\\
				P-GLMM&0.041&10.52&0.00&0.00&0.77&0.23&0.062&495.720&0.00&0.00&0.89&0.11\\
				P-GPLMM&0.036&10.70&0.00&0.00&0.93&0.07&0.038&496.250&0.00&0.00&0.92&0.08\\
				\\
				&\multicolumn{6}{c}{$(n,p)=(150,15)$}&\multicolumn{6}{c}{$(n,p)=(200,2000)$}\\
				&MSE& C(12)&I(0)&Under-fit&Correct-fit&Over-fit
				&MSE& C(1997)&I(0)&Under-fit&Correct-fit&Over-fit\\
				\cmidrule(r){2-7}\cmidrule(r){8-13}
				GPLMM&0.060&0.26&0.00&0.00&0.00&1.00&125.406&1137.62&0.00&0.00&0.00&1.00 \\
				P-GLMM&0.044&11.25& 0.00& 0.00&0.92&0.08&0.018&1996.89&0.00&0.00&0.30&0.700\\
				P-GPLMM&0.045&11.87&0.00&0.00&0.96&0.04&0.018&1996.93&0.00&0.00&0.54&0.46\\
				\\
				\hline
				\noalign{\hrule height 1pt}
			\end{tabular}
		 \end{adjustwidth}
\end{table}

\begin{table}[!h]
		\begin{center}
			\caption{Estimation results for Poisson response: performance of the P-GSMM with the cases of $p_n<n$ and $p_n>>n$. Bias: absolute value of the empirical bias; SD1: estimated standard deviation using the sandwich variance estimator; SD2: sample standard deviation; CP: denotes the empirical coverage probability of the 95\% confidence interval.}\label{tab2}
			\begin{tabular}{ccccccccccc}\hline
				\noalign{\hrule height 1pt}
				\multicolumn{5}{c}{case $p_n<n$}&&\multicolumn{5}{c}{case $p_n>>n$}\\
				\cmidrule(r){1-5}\cmidrule(r){7-11}
				$(n,p_n)$&& $\beta_1$&$\beta_2$&$\beta_3$&&$(n,p_n)$&& $\beta_1$&$\beta_2$&$\beta_3$\\
				\cmidrule(r){1-5}\cmidrule(r){7-11}
				&Bias&0.047& 0.096& 0.069&&&Bias&0.147& 0.344& 0.365 \\
				$(50,11)$&SD1&0.092& 0.085& 0.113&&$(30,100)$&SD1&0.084& 0.071& 0.094 \\
				&SD2&0.097& 0.094& 0.097&&&SD2&0.162& 0.178& 0.148 \\
				&CP&0.96 &0.95& 0.92&&&CP&0.95& 0.96& 0.97\\
				\cmidrule(r){1-5}\cmidrule(r){7-11}
				&Bias&0.076& 0.103 &0.053&&&Bias&0.078& 0.061& 0.017\\
				$(100,14)$&SD1&0.071& 0.067& 0.084&&$(100,500)$&SD1&0.049& 0.044& 0.066\\
				&SD2&0.072& 0.067& 0.078&&&SD2&0.070& 0.072& 0.086\\
				&CP&0.96& 0.95& 0.96&&&CP&0.94& 0.95& 0.96\\
				\cmidrule(r){1-5}\cmidrule(r){7-11}
				&Bias&0.101 &0.124& 0.099&&&Bias&0.049& 0.072& 0.028 \\
				$(150,15)$&SD1&0.060& 0.059& 0.070&&$(200,2000)$&SD1&0.039& 0.040& 0.054\\
				&SD2&0.052& 0.059& 0.059&&&SD2&0.051& 0.052& 0.058\\
				&CP&0.97& 0.96& 0.95&&&CP&0.94& 0.92& 0.94\\
				\hline
				\noalign{\hrule height 1pt}
			\end{tabular}
		\end{center}
\end{table}

\begin{figure}[!h]
			 \begin{adjustwidth}{-1cm}{}
		\includegraphics[width=18cm, height=10cm]{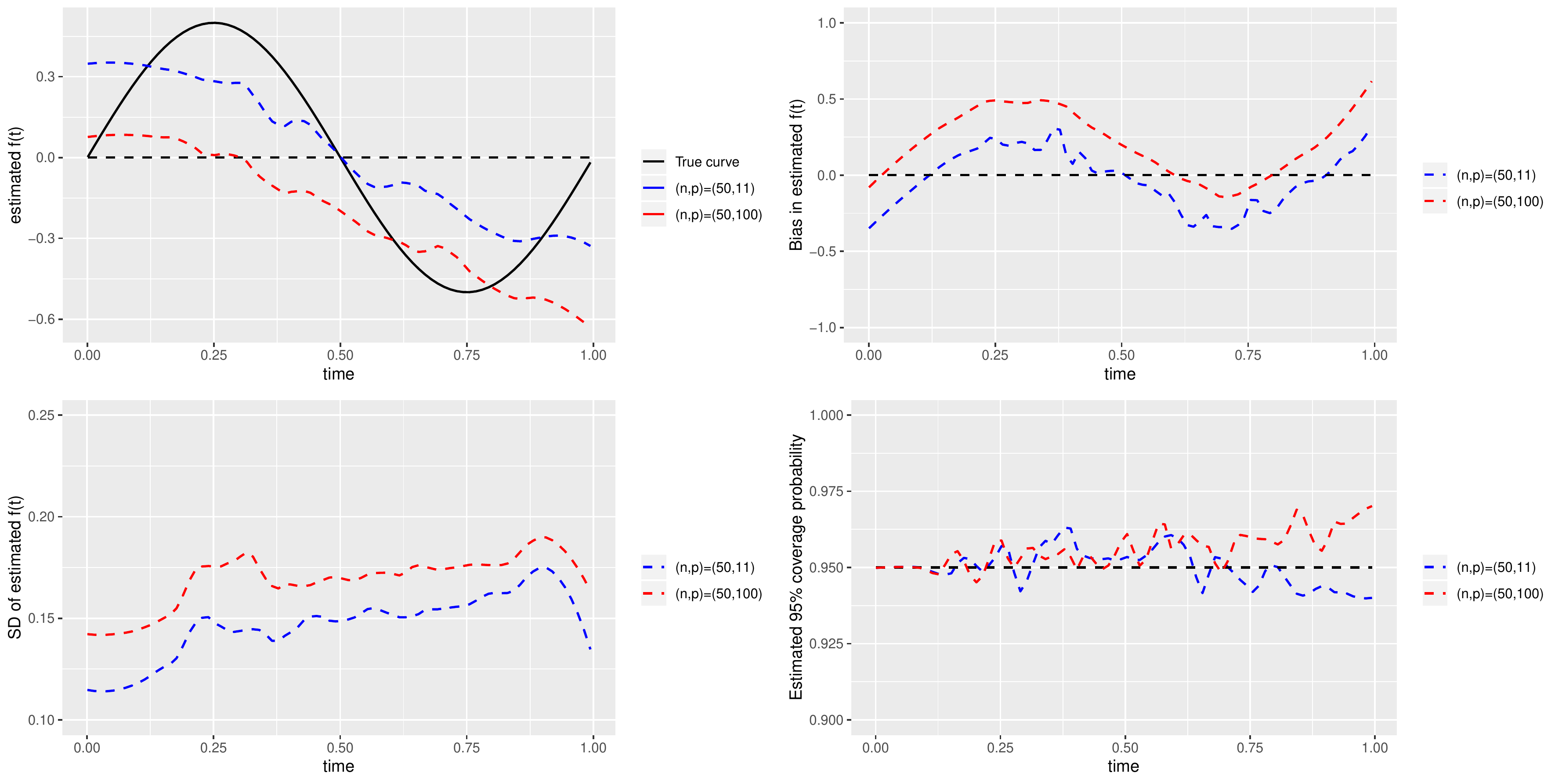}
		\caption{\small Plots for estimated $f(t)$ in the $p_n<n$ and $p_n>>n$ cases (n=50) based on 100 samples. Plots top-left and top-right show the averaged fit and pointwise bias; plot bottom-left shows the standard deviation; and plot bottom-right plots the averaged coverage probability rates for 95\% confidence intervals.}\label{fig3}
					 \end{adjustwidth}
\end{figure}

\begin{figure}[!h]
				 \begin{adjustwidth}{-1cm}{}
		\includegraphics[width=18cm, height=10cm]{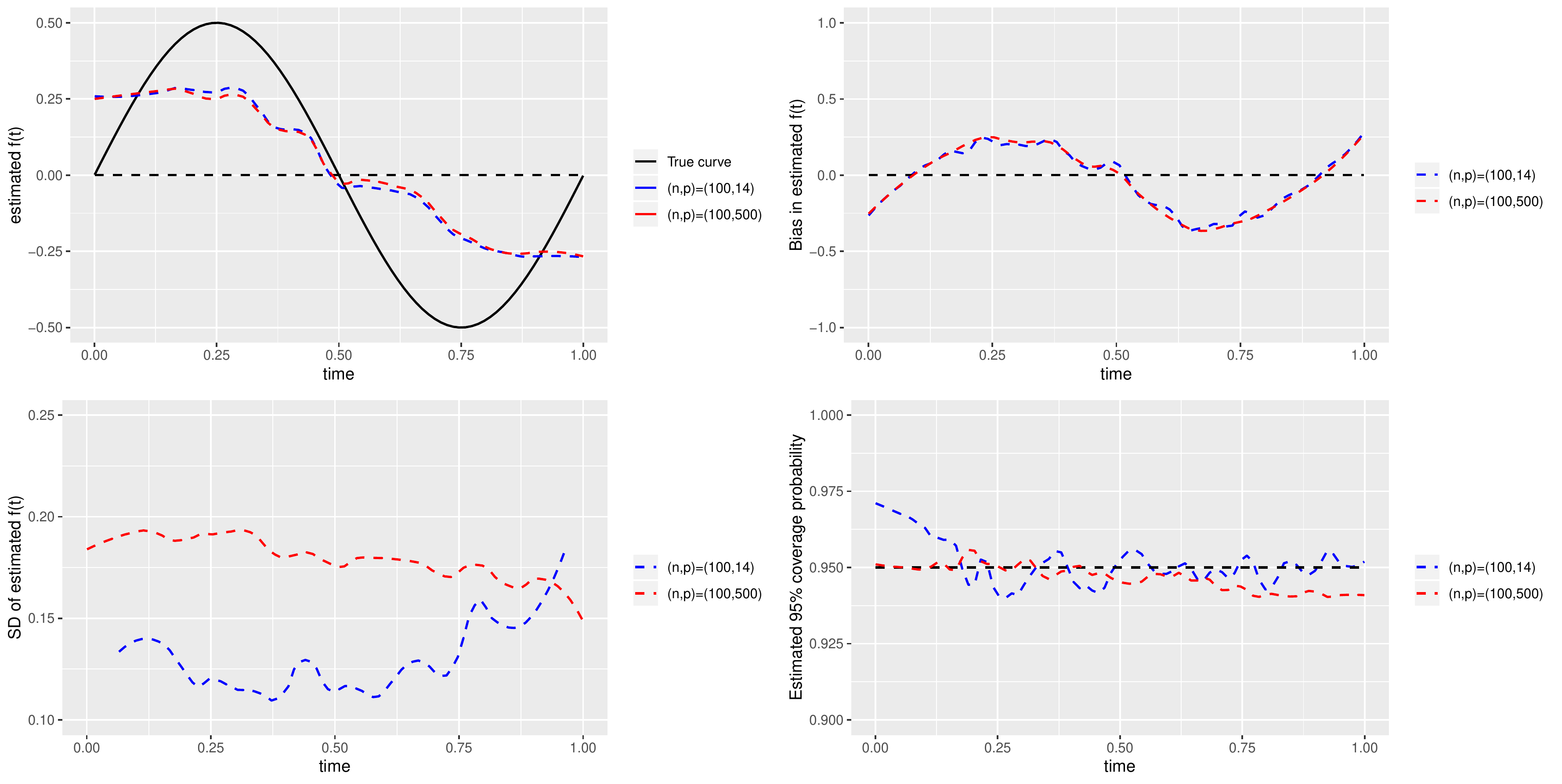}
		\caption{\small Plots for estimated $f(t)$ in the $p_n<n$ and $p_n>>n$ cases (n=100) based on 100 samples. Plots top-left and top-right show the averaged fit and pointwise bias; plot bottom-left shows the standard deviation; and plot bottom-right plots the averaged coverage probability rates for 95\% confidence intervals.}\label{fig4}
					 \end{adjustwidth}
\end{figure}

For the proposed method, the estimated baseline function $f(t)$ is also evaluated through visualization. We plot and compare the estimated
$f(t)$ and pointwise biases, for the cases of $p_n<n$ and $p_n>>n$ by two sample size $n=50$ and 100. We also plot the pointwise standard deviations (calculated from the sandwich variance formula), and
coverage probability of 95\% confidence intervals. Figures \ref{fig3} shows that for the $p_n<n$ case, our approach yields
smaller overall biases and standard deviations than the $p_n>>n$ case. Also, it can be seen that the empirical coverage probability for $f(t)$ is close to 95\% for two cases.
Figure \ref{fig4} depicts the results for $n=100$. As shown, larger sample size modified the biases and the differences between two cases. Nevertheless, the case of $p_n<n$ has smaller standard deviation.

\subsection{Real data analyses}
\subsubsection{AIDS data}
\begin{figure}[!h]
					 \begin{adjustwidth}{-1cm}{}
		\includegraphics[width=18cm, height=10cm]{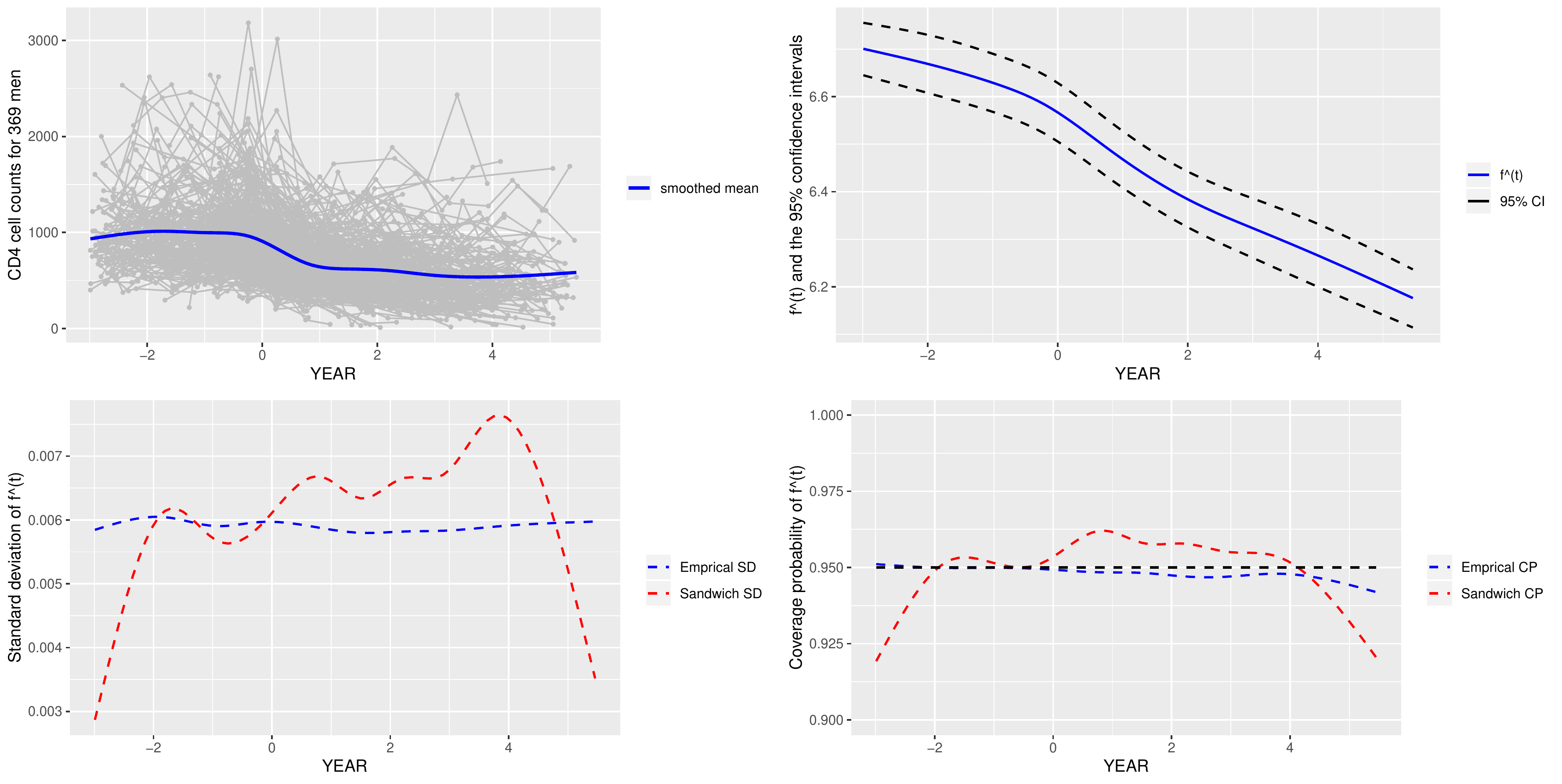}
		\caption{\small Plots for estimated $f(t)$ for AIDS data based on P-GSMM. Plot top-left shows the trajectories plot for CD4 data. Observed evolution (in gray) of CD4 cell counts for 369 men against time (in YEAR). Solid (in thick blue) line show the smoothed mean profile of men. Plot top-right shows the estimated baseline function $ f(t) $ (in thick blue) in the selected model of P-GSMM and the 95\% confidence interval (dashed line) corresponding to the robust confidence interval. Plots bottom-left and bottom-right respectively, show the standard deviation and coverage probability rates for 95\% confidence intervals based on empirical variance and sandwich formula.}\label{fig1}
					 \end{adjustwidth}
\end{figure}

In this section, to illustrate our method, we considered the longitudinal CD4 cell count data among HIV seroconverters. This dataset contains 2376 observations of CD4 cell counts on 369 men infected with the HIV virus; see \citep{Zeger:Diggle:1994} for a detailed description of this dataset. Figure \ref{fig1} (top-left) display the trajectories of 369 men for exploring the evolution of CD4 cell counts.
The first objective of this analysis is to characterize the population average time course of CD4 decay while accounting for the following additional predictor variables including AGE, SMOKE (smoking status measured by packs of cigarettes), DRUG (yes, 1; no, 0), SEXP (number of sex partners), DEPRESSION as measured by the CESD scale (larger values indicate increased depressive symptoms) and YEAR (the effect of time since seroconversion). Since there seems to exist a positive correlation among responses from the same patient, we need to incorporate a correlation structure into the estimation scheme. \cite{Zeger:Diggle:1994} found that the compound symmetry covariance matrix fitted the data reasonably well. This data analysed by many authors such as \cite{Wang:et:al:2005},\cite{Huang:et:al:2007} and \cite{Ma:et:al:2013}.

Their analysis was conducted on square root transformed CD4 numbers whose distribution is more nearly Gaussian. 
In our analysis, we fit the data using an GSMM, without transforming the CD4 by adopting the Poisson regression. To take advantage of flexibility of partially linear models, we let YEAR be modeled nonparametrically, the remaining parametrically. It is of interest to examine whether there are any interaction effects between the parametric covariates, so we included all these interactions in the parametric part. We further applied the proposed approach to select significant variables. We used the SCAD penalty, and the tuning parameter $\lambda = 0.45$.
To compare the performance of our proposed method (P-SMM) with other two existing scenarios, including the unpenalized GSMM, and the penalized GLMM (P-GLMM), we use the standard errors (SE) were all calculated using the sandwich method. To best identify a model supported by the data, we adopt the Akaike information criterion (AIC; \cite{Akaike:1973}) and the Bayesian information criterion (BIC; \cite{Schwarz:1978}). They are defined as
\begin{eqnarray}
\hbox{AIC}=2m-2\ell_{\max} , \qquad \qquad \hbox{BIC}=m\log n-2\ell_{\max}
\end{eqnarray}
where $\ell_{\max}$ is the maximized log-likelihood value, $m$ is the number of free parameters in the model. Table \ref{tab data} presents the summary of the fitting results including the values of standard errors, together with $\ell_{\max}$, AIC, and BIC under the three models. 

\begin{table}[t!]
		\caption{Summary of parameter estimates along with standard errors (in parentheses) under the three fitted models for the AIDS data.}
	\label{tab data}
	\begin{tabular}{cccc}\hline
		\noalign{\hrule height 1pt}\\
		& GSMM&P-GLMM&P-GSMM\\
		Variabeles& $\widehat{\beta}$(SE)&$\widehat{\beta}$(SE)&$\widehat{\beta}$(SE)\\		
		\cmidrule(r){1-4}
		$AGE$&~0.073~(0.039)&-0.092~(0.051)&~0~(0)\\
		$ SMOKE $&~0.188~(0.179)&~0.888~(0.192)&~0.079~(0.045)\\
		$ DRUG $&~0.130~(0.143)&~6.068(0.125)&~0.142~(0.074)\\
		$ SEXP $&-0.049~(0.031)&~0.672~(0.030)&~0.017~(0.012)\\
		$ CESD $&-0.001~(0.011)&~0~(0)&0~(0)\\
		$ AGE*SMOKE $&~0.002~(0.014)&~~0.014~(0.004)&~0~(0)\\
		$ AGE*DRUG $&-0.034~(0.024)&~0.032~(0.035)&~0~(0)\\
		$ AGE*SEXP $&-0.009~(0.003)&~0~(0)&~0~(0)\\
		$ AGE*CESD $&~0.001~(0.002)&~0~(0)&~0~(0)\\
		$ SMOKE*DRUG $&~0.009~(0.054)&-0.584~(0.150)&-0.014~(0.038)\\
		$ SMOKE*SEXP $&-0.010~(0.012)&-0.034~(0.010)&~0~(0)\\
		$ SMOKE*CESD $&-0.006~(0.009)&~0~(0)&~0~(0)\\
		$ DRUG*SEXP $&-0.025~(0.019)&-0.598~(0.041)&-0.022~(0.012)\\
		$ DRUG*CESD $&~0.006~(0.006)&~0~(0)&~0~(0)\\
		$ SEXP*CESD $&~0.001~(0.003)&~0~(0)&~0~(0)\\
		\cmidrule(r){1-4}
		$ \ell_{\max} $&~~8463007&~~7529158&~~8624429  \\
		AIC&-16925983&-15058286&-17248827\\
		BIC&-16925924&-15058228&-17248769\\
		\hline
		\noalign{\hrule height 1pt}
	\end{tabular}
\end{table}
Judging from Table \ref{tab data}, the P-GSMM tends to exhibit slightly standard errors compared to GSMM and P-GLMM, nevertheless this difference is not more dramatic. Meanwhile, the values of AIC, BIC of our proposed model are smaller than those for the other two competing models, revealing that the P-GSMM can provide better fitting performance. Under P-GSMM, SMOKE, DRUGS, SEXP, $SOMKE*DRUG$ and $DRUG*SEXP$ are identifies as significant covariates. One notes some slight selection difference when P-GLMM is used, which suggests that $ AGE*SMOKE $, $ AGE*DRUG $, and $ SMOKE*SEXP $ may also be significant. We also find some significant interactions among some covariates which may be ignored by \cite{Wang:et:al:2005} and \cite{Huang:et:al:2007}. The Results for nonparametric curve estimates using the P-GSMM estimators are plotted in Figure \ref{fig1} for $YEAR$. It shows
the estimated nonparametric function $f(t)$, its 95\% pointwise
confidence bands, standard deviation, and 95\% coverage probability given by the empirical and sandwich formula variance. We can see that the baseline function $f(t)$ has decreasing effect as time passing.
Therefore, one can see that it is more reasonable to put it as a nonparametric component.
We notice the disparity between the empirical and the sandwich formula standard deviation in the boundary positions and the sandwich formula standard deviations are smaller in which case the coverage probability recede from 95\%.

\subsubsection{Yeast cell-cycle gene expression data}
A yeast cell-cycle gene expression data collected in the CDC15 experiment of \cite{Spellman:et:al:1998} where genome-wide mRNA levels of 6178 yeast ORFs (abbreviation for open reading frames, which are DNA sequences that can determine which amino acids will be encoded by a gene) at 7 minute intervals for 119 minutes, which covers two cell-cycle periods for a total of 18 time points, measured. The cell cycle is a tightly regulated life process where cells grow, replicate their DNA, segregate their chromosomes, and divide into as many daughter cells as the environment allows. The cell-cycle process is commonly divided into M/G1-G1-S-G2-M stages. Refer to \cite{Wang:et:al:2012}, for more detailed description of this data set.

Transcription factors (TFs) have been observed to play critical roles in gene expression regulation. A TF (sometimes called a sequence-specific DNA-binding factor) is a protein that binds to specific DNA sequences, thereby controlling the flow (or transcription) of genetic information from DNA to mRNA. To better understand the phenomenon underlying cell-cycle process, it is important to identify TFs that regulate the gene expression levels of cell cycle-regulated genes. It is not clear where these TFs regulate all cell cycle genes, however.

We applied our methods to identify the key TFs. The dataset that we use present a subset of 283 cell-cycled-regularized genes observed over 4 time points at G1 stage. The response variable $Y_{ij}$ is the log-transformed gene expression level of gene $i$ measured at time point $j$, for $i = 1,\ldots, 283$ and $t = 1,\ldots,4$.  We use the following semiparametric mixed model
\begin{eqnarray*}
	y_{ij}= \sum_{k=1}^{96} x_{ij}^{(k)}+f(t_{ij})+b_i,
\end{eqnarray*}
where the covariates $x_{ij}^{(k)}$ , $k = 1,\ldots, 96$, is the matching score of the binding probability of the $k$th TF on the promoter region of the $i$th gene. The binding probability is computed using a mixture modeling approach based on data from a ChIP binding experiment; see \cite{Wang:et:al:2007} for details. Covariates $x_{ij}^{(k)}$ is standardized to have mean zero and variance 1. $t_{ij}$ denotes time, $f(t_{ij})$ models the nonparametric time effect, and $b_i$ is the random intercept. 
Our goal is to identify the TFs that might be related to the expression patterns of these 283 cellcycle–regulated genes. Therefore we apply a penalization procedure by the proposal P-GSMM and also by ignoring the nonparametric component $f(t_{ij})$ using P-GLMM.
\begin{table}[t!]
			 \begin{adjustwidth}{-1cm}{}
			\caption{Summary of parameter estimates along with standard errors (in parentheses) under the P-GSMM and P-GLMM for the Yeast Cell-Cycle Gene Expression data.}\label{tab data 2}
			\begin{tabular}{cccccc}\hline
				\noalign{\hrule height 1pt}\\
				&P-GLMM&P-GSMM&&P-GLMM&P-GSMM\\
				Variabeles&$\widehat{\beta}$(SE)&$\widehat{\beta}$(SE)&
				Continue of variables&$\widehat{\beta}$(SE)&$\widehat{\beta}$(SE)\\	
				\cmidrule(r){1-3}\cmidrule(r){4-6}
				$ARG81$&~0.022~(0.019)&0(0)&$PHD1$&0.065~(0.027)&-0.019~(0.006)\\
				$DOT6$&~0.018~(0.017)&0(0)&$RAP1$&0.053~(0.027)&0~(0)\\
				$FKH1$&0~(0)&~0.003~(0.005)&$RGM1$&0~(0)&-0.022~(0.013)\\
				$FKH2$&0~(0)&~0.166~(0.008)&$RLM1$&0~(0)&-0.002~(0.004)\\
				$GAT1$&-0.003~(0.007)&0~(0)&$RME1$&~0.072~(0.028)&0~(0)\\
				$GAT3$&~0.012~(0.014)&-0.0223~(0.012)&$SMP1$&~0.045~(0.024)&-0.015~(0.006)\\
				$MBP1$&~0.147~(0.035)&-0.1477~(0.007)&$STB1$&0~(0)&-0.008~(0.005)\\
				$MIG1$&-0.003~(0.007)&0~(0)&$STP1$&~0.002~(0.005)&0~(0)\\
				$MSN4$&~0.060~(0.027)&-0.008~(0.006)&$SWI4$&~0.076~(0.030)&-0.007~(0.006)\\
				$NDD1$&0~(0)&~0.084~(0.008)&$SWI6$&~0.1151~(0.034)&-0.020~(0.007)\\
				$PDR1$&~0.0228~(0.017)&0~(0)&$YAP5$&~0.007~(0.011)&0~(0)\\
				\cmidrule(r){1-6}
				$ \ell_{\max} $&-5.71$\times 10^{14}$&-1.58$\times 10^{14}$\\
				AIC&1.14$\times 10^{15}$&3.17$\times 10^{14}$\\
				BIC&1.14$\times 10^{15}$&3.17$\times 10^{14}$\\
				\hline
				\noalign{\hrule height 1pt}
			\end{tabular}
			 \end{adjustwidth}
\end{table}

Table \ref{tab data 2} summarizes the TFs identified when p-GSMM and p-GLMM are adopted. Our analysis reveals that a total of 13 and 16 TFs related to yeast cell-cycle processes are identified respectively by the P-GSMM and P-GLMM. The sets of TFs selected at different methods have only small overlaps. These common Tfs are GAT3, MBP1, MSN4, PHD1, SMP1, SWI4, and SWI6. For stage G1, MBP1, SWI4, and SWI6 are three TFs that have been proved important in the aforementioned biological experiments and they have been selected by the two methods. However, model selection criteria, including the values of standard errors, together with $\ell_{\max}$, AIC, and BIC confirm the superiority of our proposed model. 

\section{Conclusions}\label{sec5}
We developed a general methodology for simultaneously selecting variables and estimating the unknown components in the semiparametric mixed-effects model for non Gaussian longitudinal data when the number of parameters diverges with the sample size. Penalized estimating equation technique involves the specification of the posterior distribution of the random effects, which cannot be evaluated in a closed form, and we used a Metropolis algorithm, which does not require this specification. We further investigated some asymptotic properties of the estimates.
To investigate the performance of our approach, we compared it with the unpenalized generalized semiparametric mixed-effects model and penalized generalized linear mixed-effects model throw a simulation study and the analysis of two data sets. Results showed that the proposed model outperforms the penalized generalized linear mixed-effects counterparts on the provision of likelihood-based model selection and estimation. In addition, we found the estimation is more efficient when the partially part is taken into consideration. The results are consistent in both cases of $p_n<n$ and $p_n>>n$.
\vskip 14pt
\section*{Supplementary Materials}

The regularity conditions (A.1)-(A.8), proofs of the main results, and some instrumental lemmas are provided in a separate supplementary file.
\par




\end{document}